\documentclass[twoside,twocolumn,9pt]{article}
\usepackage{extsizes}
\usepackage[super,sort&compress,comma]{natbib} 
\usepackage[version=3]{mhchem}
\usepackage[left=1.5cm, right=1.5cm, top=1.785cm, bottom=2.0cm]{geometry}
\usepackage{balance}
\usepackage{mathptmx}
\usepackage{sectsty}
\usepackage{graphicx}

\usepackage{lastpage}
\usepackage{bm}
\usepackage[format=plain,justification=justified,singlelinecheck=false,font={stretch=1.125,small,sf},labelfont=bf,labelsep=space]{caption}
\usepackage{float}
\usepackage{fancyhdr}
\usepackage{fnpos}
\usepackage[english]{babel}
\addto{\captionsenglish}{%
  
}
\usepackage{array}
\usepackage{droidsans}
\usepackage{charter}
\usepackage[T1]{fontenc}
\usepackage[usenames,dvipsnames]{xcolor}
\usepackage{setspace}
\usepackage[compact]{titlesec}
\usepackage{hyperref}

\usepackage{epstopdf}

\definecolor{cream}{RGB}{222,217,201}

\begin{document}

\pagestyle{fancy}
\thispagestyle{plain}
\fancypagestyle{plain}{
\renewcommand{\headrulewidth}{0pt}
}

\makeFNbottom
\makeatletter
\renewcommand\LARGE{\@setfontsize\LARGE{15pt}{17}}
\renewcommand\Large{\@setfontsize\Large{12pt}{14}}
\renewcommand\large{\@setfontsize\large{10pt}{12}}
\renewcommand\footnotesize{\@setfontsize\footnotesize{7pt}{10}}
\makeatother

\renewcommand{\thefootnote}{\fnsymbol{footnote}}
\renewcommand\footnoterule{\vspace*{1pt}%
\color{cream}\hrule width 3.5in height 0.4pt \color{black}\vspace*{5pt}} 
\setcounter{secnumdepth}{5}

\makeatletter 
\renewcommand\@biblabel[1]{#1}            
\renewcommand\@makefntext[1]%
{\noindent\makebox[0pt][r]{\@thefnmark\,}#1}
\makeatother 
\renewcommand{\figurename}{\small{Fig.}~}
\sectionfont{\sffamily\Large}
\subsectionfont{\normalsize}
\subsubsectionfont{\bf}
\setstretch{1.125} 
\setlength{\skip\footins}{0.8cm}
\setlength{\footnotesep}{0.25cm}
\setlength{\jot}{10pt}
\titlespacing*{\section}{0pt}{4pt}{4pt}
\titlespacing*{\subsection}{0pt}{15pt}{1pt}

\newcommand{\Citation}{\textcolor{red}{\texttt{Citation required!}}}
\newcommand{\enote}[1]{\textcolor{blue}{\texttt{Editor: #1}}}
\newcommand{\Continue}{\textcolor{red}{\textbf{\texttt{CONTINUE HERE!}}}}
\newcommand*\mycommand[1]{\texttt{\emph{#1}}}
\newcommand{\nh}[1]{\textcolor{blue}{NH: #1}}

\newcommand{\rjm}[1]{\textcolor{purple}{RJM: #1}}

\newcommand{\clb}[1]{\textcolor{orange}{CLB: #1}}

\newcommand{\qq}{\mathbf{q}}
\newcommand{\kk}{\mathbf{k}}

\makeatletter 
\renewcommand\@biblabel[1]{#1}            
\renewcommand\@makefntext[1]%
{\noindent\makebox[0pt][r]{\@thefnmark\,}#1}
\makeatother 
\renewcommand{\figurename}{\small{Fig.}~}
\sectionfont{\sffamily\Large}
\subsectionfont{\normalsize}
\subsubsectionfont{\bf}
\setstretch{1.125} 
\setlength{\skip\footins}{0.8cm}
\setlength{\footnotesep}{0.25cm}
\setlength{\jot}{10pt}
\titlespacing*{\section}{0pt}{4pt}{4pt}
\titlespacing*{\subsection}{0pt}{15pt}{1pt}

\fancyfoot{}
\fancyfoot[LO,RE]{\vspace{-7.1pt}\includegraphics[height=9pt]{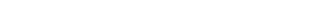}}
\fancyfoot[CO]{\vspace{-7.1pt}\hspace{11.9cm}\includegraphics{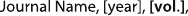}}
\fancyfoot[CE]{\vspace{-7.2pt}\hspace{-13.2cm}\includegraphics{head_foot/RF}}
\fancyfoot[RO]{\footnotesize{\sffamily{1--\pageref{LastPage} ~\textbar  \hspace{2pt}\thepage}}}
\fancyfoot[LE]{\footnotesize{\sffamily{\thepage~\textbar\hspace{4.65cm} 1--\pageref{LastPage}}}}
\fancyhead{}
\renewcommand{\headrulewidth}{0pt} 
\renewcommand{\footrulewidth}{0pt}
\setlength{\arrayrulewidth}{1pt}
\setlength{\columnsep}{6.5mm}
\setlength\bibsep{1pt}

\makeatletter 
\newlength{\figrulesep} 
\setlength{\figrulesep}{0.5\textfloatsep} 

\newcommand{\topfigrule}{\vspace*{-1pt}%
\noindent{\color{cream}\rule[-\figrulesep]{\columnwidth}{1.5pt}} }

\newcommand{\botfigrule}{\vspace*{-2pt}%
\noindent{\color{cream}\rule[\figrulesep]{\columnwidth}{1.5pt}} }

\newcommand{\dblfigrule}{\vspace*{-1pt}%
\noindent{\color{cream}\rule[-\figrulesep]{\textwidth}{1.5pt}} }

\makeatother

\twocolumn[
  \begin{@twocolumnfalse}
{\includegraphics[height=30pt]{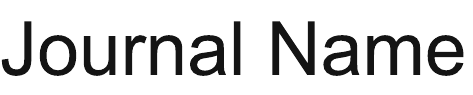}\hfill\raisebox{0pt}[0pt][0pt]{\includegraphics[height=55pt]{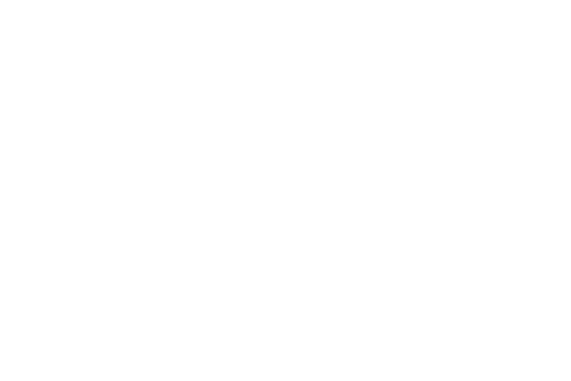}}\\[1ex]
\includegraphics[width=18.5cm]{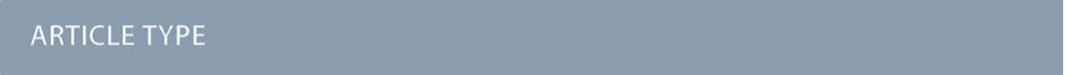}}\par
\vspace{1em}
\sffamily
\begin{tabular}{m{4.5cm} p{13.5cm} }

\includegraphics{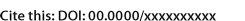} & \noindent\LARGE{\textbf{Mode selectivity in electron promoted vibrational relaxation of chemisorbed hydrogen on molybdenum and tungsten surfaces.$^\dag$}} \\
 & \vspace{0.3cm} \\

 & \noindent\large{Nils Hertl\textit{$^{a,b,*}$}, Connor L. Box\textit{$^{b}$}, Reinhard J. Maurer\textit{$^{a, b, c,*}$}} \\

\includegraphics{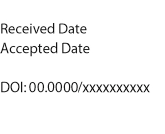} & \noindent\normalsize{ 
Electron-phonon coupling in atoms and molecules adsorbed at metal surfaces gives rise to finite vibrational linewidths in infrared or electron energy loss spectra. When it is the dominant contribution to the vibrational lifetime, it manifests itself in the form of a Fano line shape. Here, we report the linewidths of vibrational modes of chemisorbed hydrogen on the (100) and (110) surfaces of molybdenum and tungsten calculated from first-order time-dependent perturbation theory. For those modes with a Fano line shape, our results are in good agreement with the experiment. We further observe that the coupling strength between vibrations and electrons depends on the nature of the mode: for Lorentzian-shaped peaks, the experimental linewidths are always larger than those predicted based on pure electron-phonon coupling. The calculated linewidths exhibit a strong coverage dependence, decreasing towards higher coverages. This finding has important implications for nonadiabatic energy dissipation in hydrogen dynamics at metal surfaces. 
While electron–hole pair excitation is the dominant energy-transfer mechanism between hydrogen and pristine metal surfaces, other channels for energy dissipation, such as adsorbate–adsorbate interactions, may become more significant on metal surfaces densely covered with hydrogen. }

\end{tabular}

 \end{@twocolumnfalse} \vspace{0.6cm}

  ]

\renewcommand*\rmdefault{bch}\normalfont\upshape
\rmfamily
\section*{}
\vspace{-1cm}

\footnotetext{\textit{$^{a}$ Department of Physics, University of Warwick, Gibbet Hill Road, Coventry CV4 7AL, United Kingdom. }}
\footnotetext{\textit{$^{b}$Department of Chemistry, University of Warwick, Gibbet Hill Road, Coventry CV4 7AL, United Kingdom.}}
\footnotetext{\textit{$^{c}$Faculty of Physics, University of Vienna, Kolingasse 14{--}16, 1090 Vienna, Austria}}
\footnotetext{\textit{$^{*}$ Corresponding Authors: nils.hertl@warwick.ac.uk, reinhard.maurer@univie.ac.at}}

\footnotetext{\dag~Electronic Supplementary Information (ESI) available: Visualised comparison of PES with DFT-based input data for both surface facets. See DOI: 00.0000/00000000.}

\footnotetext{\ddag~Additional footnotes to the title and authors can be included \textit{e.g.}\ `Present address:' or `These authors contributed equally to this work' as above using the symbols: \ddag, \textsection, and \P. Please place the appropriate symbol next to the author's name and include a \texttt{\textbackslash footnotetext} entry in the the correct place in the list.}

\vspace{1cm}
\section*{Introduction}

Adsorption of hydrogen can induce significant changes in the electronic structure, leading to enhanced electron-phonon coupling in two-dimensional materials,\cite{bekaert2019, Yan2022, soskic2024} or at metal surfaces such as tungsten.\cite{Rotenberg1998, Rotenberg2000, Rotenberg2002} Electron-phonon coupling is pivotal to driving phenomena in hydrogen-metal systems ranging from the Jahn-Teller effect\cite{Jahn1937, bersuker2020}, superconductivity\cite{duan2017, semenok2020} to vibrational relaxation at surfaces.\cite{Persson1982, gadzuk1984, tully2000b} Hydrogen adsorption is also a crucial step  in the formation of metal hydrides\cite{bloch1997} and in heterogeneously catalysed reactions at surfaces such as the Haber-Bosch process\cite{humphreys2021} or the Fischer-Tropsch process.\cite{dry2002} Moreover, adsorption of atomic hydrogen on tungsten surfaces is an elementary step that contributes  to the macroscopic recycling coefficient of plasma fuel in a fusion reactor.\cite{causey2002} Yet, incoming hydrogen will only adsorb when the excess energy is efficiently dissipated. At metal surfaces, the dissipation of the excess energy released during H$_2$ bond dissociation is governed to a large extent by its interaction with the electrons of the substrate. In this scenario, electron-phonon coupling enables energy dissipation via excitation of electron-hole pairs (EHPs) triggered by the motion of the adsorbed atom.~\cite{Persson1982, gadzuk1984, tully2000b} Other dissipation mechanisms like the coupling of the hydrogen atom motion to the surface phonons of the substrate are considered to be inefficient due to the large mass mismatch between the adsorbed hydrogen's mass and the atomic mass of the substrate.\cite{norskov1979, grimmelmann80}

Vibrational spectra of atoms and molecules on surfaces can reveal much about the nature and effectiveness of energy exchange between adsorbate and substrate, because the dissipation rate contributes intrinsically to the overall linewidth of individual spectral signals. However, isolating the different broadening contributions is challenging, as inhomogeneous broadening, pure dephasing, and instrumental (Gaussian) broadening all contribute to the overall linewidth.\cite{gadzuk1984, guyot1995surface} Sum-frequency generation (SFG) experiments provide a means to directly measure the depopulation time and thus decipher the kinetics of energy relaxation from the broadening caused by pure dephasing.\cite{guyot1995surface} Yet, the depopulation time, $T_1$, is the inverse of the relaxation rate of \textit{all} mechanisms that dampen the vibration. Thus, experiments have often been performed in conjunction with theoretical support, allowing for an estimate of the effectiveness of electron-phonon-mediated dissipation in comparison to phonon-phonon-mediated dissipation.\cite{Ryberg1981, Bpersson1982, morin1992, head-gordon1992, tully93, Persson1989}

From a theoretical standpoint, EHP-mediated vibrational relaxation is commonly described within the framework of electronic friction.\cite{dAgliano75, head-gordon95} In this framework, the coupling between the vibrations and the electrons of the substrate is assumed to be weak and can therefore be derived with time-dependent perturbation theory, which yields a Fermi's golden rule type expression for the dampening rate.\cite{Persson1982, Bpersson1982, hellsing84, head-gordon95} Recent developments have connected this formalism to full-potential Kohn--Sham density functional theory within the numeric atomic orbital software FHI-aims~\cite{aims01}, enabling efficient \textit{ab initio} calculations of mode-resolved linewidths and the corresponding friction tensors for adsorbates on metal surfaces.\cite{askerka16, maurer16, box2024}

In addition to the linewidth, the shape of the spectral line can also contain information about the broadening mechanism. Langreth showed that a vibrational signature has a line shape of the Fano-type when its relaxation is dominated by an EHP dampening mechanism.\cite{Langreth1985, Langreth1987} In fact, Fano line shapes have been observed in infrared (IR) reflection experiments on H/Mo(100)\cite{Reutt1988}, H/W(100)\cite{Chabal1985, Reutt1988}, and H/Cu(111)\cite{lamont1995}, as well as for CO/Cu(100).\cite{hirschmugl1990} Electron energy loss spectroscopy (EELS) has likewise revealed Fano-type asymmetry for H/Mo(110)\cite{Kroger1997} and H/W(110).\cite{Balden1996} More recently, asymmetric line shapes have also been reported for polyatomic molecules on coinage-metal surfaces.\cite{jakob2021, priya2024} Aside from IR and EELS experiments, Fano line shapes have also been observed in IETS measurements for chemisorbed hydrogen on Cu(100).\cite{lauhon2000} These line shapes have been characterised theoretically using the theory developed by Lorente and Persson.\cite{lorente2000a, lorente2000b, paavilainen2006} The line shape formula derived by Langreth has been used to characterise these experimental observations, but to the best of our knowledge, no first-principles characterisation of the experimental IR and EELS experiments on full monolayer coverages of H/metal surfaces for Mo and W has been performed.

In this work, we leverage the time-dependent perturbation theory framework to compute vibrational linewidths for the different modes of hydrogen and deuterium adsorbed on the (100) and (110) surface facets of molybdenum and tungsten. Our calculations reproduce the linewidths of those vibrational modes that exhibit Fano line shapes in experiment, supporting the interpretation that these modes are predominantly broadened by EHP excitations. In contrast, vibrations on the (100) surfaces that display Lorentzian profiles in IR and EELS spectra show significantly broader experimental linewidths than predicted by our calculations, indicating that additional broadening mechanisms beyond EHP-mediated relaxation contribute strongly in this case. This also allows for a realistic assessment of the extent to which EHP excitations contribute to the overall linewidth of the remaining modes. Finally, we analyse the influence of hydrogen coverage on the vibrational linewidth and electronic friction magnitude. We find that high coverages lead to markedly reduced linewidths (decreased electronic friction).  We discuss the implications of these findings for molecular dynamics with electronic friction.~\cite{head-gordon95} Essentially, electronic friction frameworks that rely in practice on the electronic density of the pristine metal\cite{li_wahnstrom92a, juaristi08} might overestimate electronically non-adiabatic effects when used to model processes at high-coverage hydrogen-metal surfaces.\cite{juaristi2017, martin-barrios2021, martin-barrios2024}

\section*{Methods}

\subsection*{Computational Details}
All calculations were performed using density functional theory (DFT) with the PBE exchange-correlation functional\cite{pbe1, pbe2} as implemented in the \texttt{FHI-aims} software package (version 20251007).\cite{aims01, aims02, aims03, aims04, aims05} Scalar-relativistic corrections were included via the atomic ZORA approach.\cite{ZORA} Our employed convergence criteria for the energy, eigenvalues, density, and forces are $10^{-6}$\,eV, $10^{-3}$\,eV, and $10^{-5}$\,e/$\text{a}_0^3$, respectively. For all structural optimisation calculations, we used the default "tight" basis set definition. For all calculations, we used a Gaussian occupation smearing width of 0.1~eV. For the lattice constant optimisation, the Brillouin zone was sampled by a $36\times36\times36$ $\boldsymbol{k}$-point mesh with sampling method proposed by Monkhorst and Pack.\cite{monkhorst76}  The optimised lattice constants for Mo and W are 3.16 and 3.18\,{\AA}, respectively, and are in good agreement with previous calculations employing the same functional\cite{hertl22, ghosh2023} as well as experimental values.\cite{wyckoff1963}

\begin{figure*}
    \centering
    \includegraphics[width=1\linewidth]{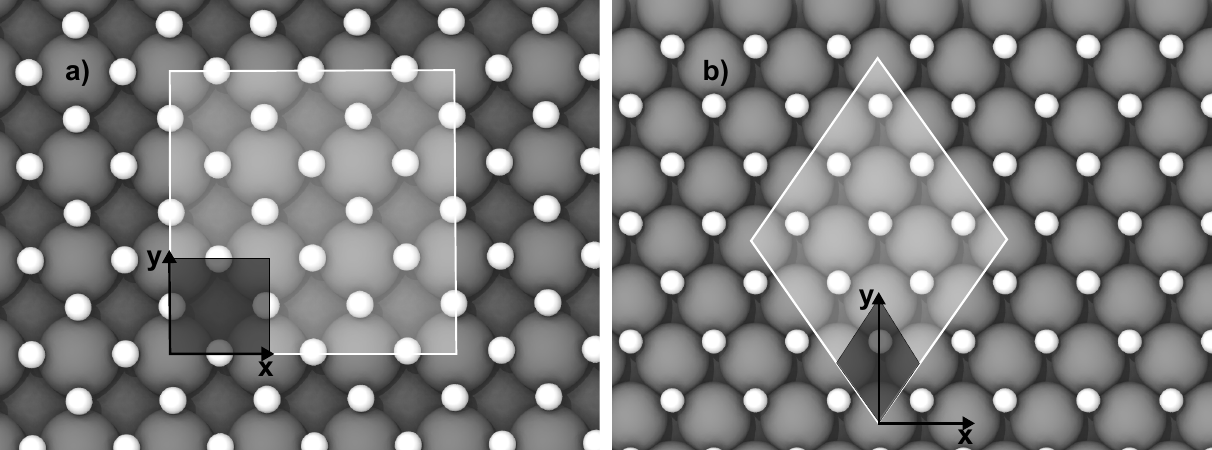}
    \caption{Orthographic view on the $p(1\times1)$H/bcc(100) surface shown in panel a) and the $p(1\times1)$H/bcc(110) surface in panel b). The black arrows define the Cartesian coordinates parallel to the surface. The black and white opaque areas define the primitive cell and a $3\times3\times1$  supercell of the H/metal surfaces, respectively.}
    \label{fig:metal_surfaces}
\end{figure*}
All H/metal surfaces were modelled as periodically repeated slabs consisting of six layers, of which the two lowest layers remained fixed in the geometry optimisation.  The forces have been minimised until the convergence criterion of $10^{-3}$\,eV/\AA{} was met. For the primitive $p(1\times1)$H/metal cells, we placed two hydrogen atoms on the bridge site in case of the (100) surface, and one hydrogen on the three-fold hollow site for the (110) surface, which yields a hydrogen coverage, $\Theta$, of 1~ML. The structures are represented by the black, opaque areas in Fig.\,\ref{fig:metal_surfaces}. For the primitive cells used in the geometry optimisation, the Brillouin zone was sampled by a $36\times36\times1$ $\boldsymbol{k}$-point mesh. Lower coverages for the H/metal(110) surfaces were realised by placing one hydrogen on $n$-supercells. To ensure the same $\boldsymbol{k}$-grid density throughout the different $n\times n \times 1$ supercells, we divided the k-point mesh by the factor of $n$. 

\subsection*{Electron-Phonon Coupling and Linewidths}
Harmonic vibrational frequencies and normal modes were calculated from the dynamical matrix, constructed with finite difference displacement of $10^{-3}$~\AA{} with respect to the relaxed atomic positions. The calculations of the frequency-dependent electron-phonon rates are based on the electronic friction formalism reported by Box \textit{et al.}\cite{Box2023}, which are implemented in \texttt{FHI-aims} version 20251007. For both the vibrational frequencies $\omega$ and rate tensor $\boldsymbol{\Lambda}(\hbar\omega)$, we omitted the degrees of freedom of the metal atoms. While this is in line with previous calculations of $\boldsymbol{\Lambda}(\hbar\omega)$ for small adsorbates on metal surfaces,\cite{maurer16, Box2023, stark2025}, it is also well justified for the vibrational modes of hydrogen due to the large mass mismatch between H and Mo and W, which leads to energetically well-separated phonon bands.\cite{Eiguren2008} Consequently, the coupling between the hydrogen phonon bands and the metal is weak.

The frequency-dependent electron-phonon relaxation rate tensor in Cartesian coordinates $\boldsymbol{\Lambda}(\hbar\omega)$ was evaluated using first-order time-dependent perturbation theory (TDPT) for phonons at the $\Gamma$-point, i.e., $\mathbf \qq=0$:
\begin{equation}
\begin{aligned}
\Lambda_{a\kappa,a'\kappa'}(\hbar\omega)
= 2\pi\hbar \sum_{ mn}\!\int_{\mathrm{BZ}}\!\frac{d\mathbf k}{\Omega_{\mathrm{BZ}}}\;
\tilde g^{a\kappa}_{mn}(\mathbf k)\,
\tilde g^{a'\kappa'}_{mn}(\mathbf k)^{*}\, \\ \cdot
\frac{f_{n\mathbf k}-f_{m\mathbf k}}{\epsilon_{m\mathbf k}-\epsilon_{n\mathbf k}}\,
\delta\!\big(\epsilon_{m\mathbf k}-\epsilon_{n\mathbf k}-\hbar\omega\big),
\end{aligned}
\label{eq:LambdaGamma}
\end{equation}
where $a,a'$ are atom indices, $\kappa,\kappa'\in\{x,y,z\}$ are Cartesian coordinates, $m,n$ are Kohn--Sham state labels, $f_{n\mathbf k}$ are Fermi occupations, and $\Omega_{\mathrm{BZ}}$ is the Brillouin-zone volume. Equation~\ref{eq:LambdaGamma} describes the first-order 
interband electronic transitions induced by atom displacement.  
The electron--phonon matrix elements 
$\tilde g^{a\kappa}_{mn}(\mathbf k)$ couple pairs of occupied and unoccupied 
Kohn--Sham states that differ in energy by $\hbar\omega$, corresponding to the 
absorption or emission of a vibrational quantum (with momentum $\qq=0$). In practice, the integral was evaluated on a $4\times4$ six-layer slab of the $p(1\times1)$\,H/metal$(hkl)$ systems using a $9\times9\times1$ of the Brillouin zone as described above, and the $\delta$-function was replaced by a normalised Gaussian of width 0.075~eV to ensure convergence of the 
frequency-resolved coupling matrix. 
For the surface structures with different hydrogen coverages, $\Theta$, which were modelled using supercells of varying lateral size, we evaluated Eq.,\ref{eq:LambdaGamma} using $\boldsymbol{k}$-point grids adjusted to maintain a constant $\boldsymbol{k}$-grid density. For each structure, the $\boldsymbol{k}$-point grid employed to compute the friction tensor was identical to the one used in the geometry optimisation of the corresponding slab, as described above.
  As previous benchmarks \cite{stark2025} for H on metals show a weak influence of the chosen basis set on the TDPT calculations, we used a light basis set, but with a dense integration grid for converged forces. The temperature in the Fermi-distribution $f_{n\mathbf k}$ is set to 100\,K, which is close to the experimental conditions.\cite{Reutt1988, Kroger1997} The factor of two in Equation~\ref{eq:LambdaGamma} arises from ignoring the explicit sum over spin channels, which is suitable for H atoms close to a metallic surface.~\cite{box2024}

Projection of $\boldsymbol{\Lambda}$ onto a mass-weighted, normal mode displacement vector of mode $\nu$, $\tilde{\mathbf u}_{\nu}$, (from the same simulation cell) yields the mode-resolved linewidth $\gamma_{\nu}$, corresponding to the full width at half maximum (FWHM), and corresponding lifetime due to electron-hole pair dissipation, $\tau_{\nu}$:
\begin{align}
\gamma_{\nu} &
= \hbar\, \tilde{\mathbf u}_{\nu}\,
\Lambda(\hbar\omega_{\nu})
\tilde{\mathbf u}_{\nu}^{\mathrm T}, \label{eq:linewidth} \\
\tau_{\nu} & =\hbar/\gamma_{\nu}.
\label{eq:linewidth_lifetime}
\end{align}
Since the phonon bands we consider in this work are monoisotopic, we can reformulate the expression on the right-hand side of Eq.\,\ref{eq:linewidth}
\begin{equation}
    \gamma_{\nu} =  \frac{\hbar}{M_\nu} \mathbf u_{\nu}\Lambda(\hbar\omega_{\nu}) \mathbf u_{\nu}^{\mathrm T},
    \label{eq:rate_isotope_dependence}
\end{equation}
where we have used the following relation: $\tilde{\mathbf u}_{\nu} =  \mathbf{u}_{\nu} /\sqrt{M_\nu}$.
From the $1/M$ dependence, it can also be inferred that the contribution from neighbouring metal atoms to the linewidth of the hydrogen vibrations are small. 
\subsection*{Identification of Coherent and Averaged Modes}

 Although the implementation operates strictly at \(\mathbf q=0\), we employed larger real-space supercells $(n_x\times n_y\times n_z)$, such that band folding provides us with access to phonon modes with $\qq \neq 0$.  As a result, a single supercell $\Gamma$-point calculation contains a dense manifold of modes representing different phase relations between primitive-cell repeats. 
 
 In a finite real-space supercell, all vibrational modes are formally
labelled as $\Gamma$-point modes of that supercell. However, these
supercell $\Gamma$-point modes do not all correspond to the true
$\mathbf q{=}0$ modes of the primitive cell. Enlarging the real-space
cell reduces the size of the Brillouin zone, such that phonons with
finite wave-vectors $\mathbf q$ in the primitive cell are
folded back to the $\Gamma$-point of the supercell. Each such
folded mode appears as a distinct eigenmode with the same local
adsorbate motion (e.g.\ wagging or stretching), but with different
phase relations between neighbouring primitive cells.

Among the supercell modes, one displacement pattern corresponds to the coherent $\qq=0$ motion, where all adsorbate images move in phase across the supercell. As the wavelength of light used is typically much larger than the lattice constant, this mode predominantly gets excited in spectroscopic experiments before it dephases into modes with finite wave-vectors $\qq$.  The $\qq=0$ mode is straightforward to identify visually in the eigenvector animations and was used to obtain the ``coherent'' linewidth \(\gamma_{0}\) reported in Table~\ref{tab:Linewidths_all_facets_et_isotopes}.

The remaining modes with the same vibrational character but different
phase relations represent finite-$\mathbf q$ vibrations of the primitive
cell that have been folded to the supercell $\Gamma$-point. Since the employed implementation operates strictly at the supercell
$\Gamma$-point, evaluating linewidths for these folded modes provides a
practical route to accessing electron--phonon dissipation for
finite-$\mathbf q$ vibrations without explicitly performing
$\mathbf q$-dependent calculations. 


To get an upper limit for the linewidths, we assume that the light-excited coherent mode $\nu$ at $\qq=0$ dephases instantaneously into a localised vibration that can be represented as a linear combination of mod $\nu$ at all possible wave-vectors $\qq$. 
As detailed by Box \textit{et al.},\cite{Box2023} the linewidth of phonon mode $\nu$ in this limit of instantaneous dephasing can be written as:
\begin{equation}
    \bar{\gamma} = \frac{1}{N_{\qq}} \sum_{\qq} \gamma_{\nu,\qq},
\end{equation}
where $N_{\qq}$ is the number of $\qq$-points.
To calculate the linewidth in this limit across the adlayer, we identified all  modes of the
same phonon band $\nu$ within a narrow frequency window (typically
$\pm 20$--$30~\text{cm}^{-1}$) and computed their individual linewidths.
The average of these linewidths, referred to as ``$\qq$-averaged''
linewidth, $\bar{\gamma}$, represents the depopulation rate of an
incoherent ensemble of locally similar vibrations.~\cite{Box2023}

Together, $\gamma_{0}$ and $\bar{\gamma}$ provide lower and upper bounds
on the first-order electron--phonon contribution to the linewidth:
$\gamma_{0}$ corresponds to the perfectly coherent limit where all
adsorbates oscillate in phase, while $\bar{\gamma}$ represents the
instantaneously dephased limit in which neighbouring adsorbates move
with arbitrary phase relations.

\section*{Results and Discussion}

\subsection*{Mode Characterisation}
\begin{figure}
    \centering
    \includegraphics[width=3.3in]{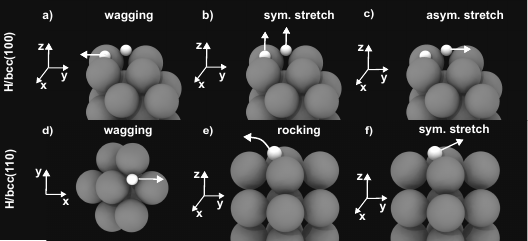}
    \caption{Adsorbate vibrational modes for the $p(1\times1)$ phases of hydrogen adsorbed on molybdenum and tungsten. Panels a){--}c) display the vibrations on the (100) facet, whereas panels d){--}f) shows the vibrations of the (110) facet. Note that the wagging mode in panel a) and the asymmetric stretch mode in panel c) have a two-fold degeneracy. }
    \label{fig:vibrations}
\end{figure}
We now discuss the modes of the hydrogen vibrations on the (100) and (110) surfaces for both molybdenum and tungsten.  The structures of the $p(1\times1)$H/metal($100$) and H/metal($110$) surfaces are shown in Figure\,\ref{fig:metal_surfaces}. The $p(1\times1)$H/bcc(100) structure for molybdenum and tungsten is constituted by two H atoms per unit cell adsorbed on the bridge sites between two metal atoms.\cite{Willis1974, Ho1978, Biswas1986} The H atoms in the $p(1\times1)$H/M(110) phase for molybdenum and tungsten are located in the centre of the triangle formed by the metal atoms, usually referred to as three-fold site.\cite{Willis1974, kwak1996, Kroger1997, Balden1996, Kohler1997}

Both $p(1\times1)$H/Mo(100) and $p(1\times1)$H/W(100) have an in-plane wagging mode, an out-of-plane symmetric stretch mode, and an in-plane asymmetric stretch mode.\cite{Willis1974, Ho1978, Zaera1986, Smith1991}  For both systems, the in-plane wagging mode is the lowest-energy mode, followed by the out-of-plane mode and the planar asymmetric stretch mode is the highest energetic mode.\cite{Reutt1988, Smith1991, Lou1990, Zaera1986}  
Both the $p(1\times1)$H/Mo(110) and $p(1\times1)$H/W(110) hydrogen vibrations can be characterised as a planar wagging mode with no normal component, a rocking mode, and a symmetric stretch mode.\cite{Balden1996, Kroger1997} The vibrations are displayed in Figure\,\ref{fig:vibrations}. The corresponding computed frequencies of these modes are in good agreement with experimental EELS data (Table\,\ref{tab:modes}).

In infrared absorption spectroscopy experiments conducted by Chabal\cite{Chabal1985} on $p(1\times1)$H/W(100), two absorption signals were observed: one at 1070\,cm$^{-1}$ with a symmetric Lorentzian line shape and one at 1270\,cm$^{-1}$ with an asymmetric line shape of the Fano type. Reutt \textit{et al.}\cite{Reutt1988} made similar observations for  $p(1\times1)$H/Mo(100): they found a symmetric signal with a Lorentzian line shape at 1016\,cm$^{-1}$ and an asymmetric signal at 1300\,cm$^{-1}$. Both studies assigned the signal with the Lorentzian line shape to the symmetric stretch mode and the signal with the Fano line shape to the overtone of the wagging mode.\cite{Chabal1985, Reutt1988} However, given that the overtone of the wagging mode and the fundamental of the asymmetric stretch mode are nearly isoenergetic for H/W(100), this interpretation was questioned.\cite{Zhang1989} With the help of angle-resolved photoelectron spectroscopy (ARPES) measurements and symmetry arguments, the Fano-type line observed in the IR experiment on H/W(100) could be unambiguously assigned to the asymmetric stretch mode.\cite{smith1990} For H/Mo(100), however, ARPES measurement remained inconclusive about the assignment of the mode and it was argued that in fact both modes may contribute to the signal observed in the IR experiments.\cite{Smith1991} Yet, with 555\,cm$^{-1}$, the frequency for the fundamental of the wagging mode measured with EELS by Zaera \textit{et al.}\cite{Zaera1986} is not half the size of the fundamental of the asymmetric stretch mode (1260\,cm$^{-1}$). In fact, the overtone would be expected to be in the area of $\sim$1100\,cm$^{-1}$. Therefore, it is more natural to assign the asymmetric signal at 1300\,cm$^{-1}$ in the IR spectrum for H/Mo(100) to the asymmetric stretch mode rather than the first overtone of the wagging mode. Asymmetric line shapes of the Fano type have also been observed EELS measurements for $p(1\times1)$H/Mo(110) which could be unambiguously assigned to the rocking mode.\cite{Kroger1997} For H/W(110) a continuum of signals starting at about 810\,cm$^{-1}$ was reported which the authors interpreted as a 2D liquid phase.\cite{Balden1996}

\begin{table}[]
\centering
\caption{Mode type and computed frequencies for the four $p(1\times1)$H/metal($hkl$) systems taken from a $(4\times4\times6)$ slab along with experimental frequencies obtained from EELS measurements. The particular mode of each H/metal system, which exhibits a Fano line shape in spectroscopic experiments, is indicated by a $\dagger$ symbol.}
\label{tab:modes}
\begin{tabular}{llcc}
\hline
\textbf{Surface facet} & \textbf{Mode} & $\omega_{\text{DFT}}$/ cm$^{-1}$  & $\omega_{\text{Lit}}$ / cm$^{-1}$ \\
\hline
        & wagging  & 566 &  555\cite{Zaera1986}\\  
Mo(100) & sym. strech & 1034 & 1025\cite{Zaera1986}  \\
        & asym. stretch$^\dagger$ & 1342 & 1260\cite{Zaera1986}  \\
\hline
        & wagging  & 665 & 645\cite{Ho1978} \\
W(100)  & sym. strech & 981 & 1048\cite{Ho1978} \\
        & asym. stretch$^\dagger$ & 1282 & 1290\cite{Ho1978} \\
\hline
        & wagging   & 785 & 742\cite{Kroger1997} \\
Mo(110) & rocking$^\dagger$  & 853  & 802 \cite{Kroger1997} \\
        & sym. stretch & 1261  & 1229 \cite{Kroger1997}\\
\hline
        & wagging   & 686  & 655\cite{Balden1996} \\
W(110)  & rocking  & 858  & 809 \cite{Balden1996}\\
        & sym. stretch & 1338 & 1300 \cite{Balden1996}\\
\hline

\end{tabular}
\end{table}

\subsection*{Vibrational Linewidths for the $\boldsymbol{p(1\times1)}$H/metal surfaces}

In this section, we discuss the linewidths due to the EHP-excitation mechanism for the three different hydrogen vibration modes of the four H/metal surfaces and compare them to experimental findings reported in Refs.\citenum{Reutt1988, Kroger1997}. 
Adsorbate vibrations with an asymmetric line shape in spectroscopic experiments are of particular interest because Fano-type line shapes are fingerprints of non-adiabatic effects in vibrational energy dissipation.\cite{Langreth1985}  
The expression for the line shape of such a phenomenon is
\begin{equation}
    L(\omega) = \frac{4\omega_\text{r}\tau  (\mu_1)^2}{\gamma} \frac{(1-xy)^2}{(1+x^2)},
    \label{eq:Fano_line}
\end{equation}
where $x=(\omega^2-\omega_\text{r}^2)/\gamma\omega$ and $y=\omega \tau$. Note that the line shape in Eq.,\ref{eq:Fano_line} is a generalisation of the Fano line shape,\cite{Langreth1985} and the connection to the original work by Fano\cite{fano1961} has been discussed by Sorbello.\cite{sorbello1985}
The constant $\omega_\text{r}$ is the renormalised adsorption frequency. 
$\gamma$ is the full-width at half maximum (FWHM) of the line shape $L(\omega)$ and is entirely caused by EHP excitations.  
$\mu_1$ is the real component of the dynamic dipole of the adsorbate-substrate system, and $\tau$ is the inverse of the tunnelling rate of electrons and holes between the adsorbate and substrate states. 
Since the tunnelling is not instantaneous, these fluctuating EHPs cause an oscillating, imaginary contribution to the dynamic dipole, which is out of phase with the electromagnetic frequency $\omega$ of the strength $\omega\tau$. 
It is this part of the dynamic dipole which gives rise to an asymmetric line shape. 
Hence, following Langreth's arguments, those vibrations which show a Fano line shape in IR and EELS experiments relax predominantly due to EHP excitations. 
Eq.\,\ref{eq:Fano_line} was in fact fitted by Reutt \textit{et al.}\cite{Reutt1988}, Chabal\cite{Chabal1985}, and Kr\"oger \textit{et al.}\cite{Kroger1997} to the measured asymmetric lines to characterise the strength of the EHP-mediated vibrational relaxation. 
This connection enables a meaningful comparison between the experimentally extracted FWHMs of the asymmetric (Fano) modes and our calculated averaged linewidths, which represent the fully dephased, EHP–induced relaxation rate of those vibrations.

The results from our calculations for the coherent vibrational linewidth, $\gamma_0$,  and the linewidth of the $\qq$-averaged mode, $\bar{\gamma}$, are given in Table~\ref{tab:Linewidths_all_facets_et_isotopes} for both hydrogen and deuterium. In all reported cases, $\gamma_0$ is smaller than the $\bar{\gamma}$ values, which is consistent with previous observations for vibrational lifetimes of $c(2\times2)$CO/Cu(100) computed with the same framework as in this work.~\cite{maurer16, Box2023} Novko \textit{et al.} further showed that the $\qq>0$ modes have larger vibrational linewidths for the internal stretch vibration of CO adsorbed on Cu(100).\cite{Novko2018, Novko2019} We make the same observations for all modes in the $(1\times1)$ hydrogen covered phases on molybdenum and tungsten surfaces. The coherent modes have linewidths below 10\,cm$^{-1}$, except for the H/Mo(100) symmetric stretch mode and the H/Mo(110) wagging mode which exhibit linewidths of 13\,cm$^{-1}$ and 12\,cm$^{-1}$, respectively. The averaged modes have linewidths up to 30\,cm$^{-1}$. 
When comparing linewidths between Mo and W for the same surface facet, the linewidths on Mo are generally larger.
For the coherent modes, $\gamma_{0}$ values on the (100) facets follow the same ordering: the linewidth of the symmetric stretch mode is the largest, followed by the asymmetric stretch mode and then the wagging mode. For the (110) surfaces, the wagging mode is the largest one for both elements, but otherwise, the order for the other modes is swapped. 
In general, $\gamma_{0}$ values are smaller and more sensitive to details of the electronic structure description, 
making systematic trends difficult to establish. 
For the averaged linewidths $\bar{\gamma}$, clearer differences emerge.  
Hydrogen vibrations on Mo(100) exhibit noticeably broader $\bar{\gamma}$ values than those on W(100), 
indicating stronger electronic coupling on the molybdenum surface.  
On the (110) facets, the averaged linewidths for molybdenum and tungsten are of comparable magnitude, except for the wagging mode, which is significantly broader for Mo(110).  
For deuterium, linewidths are typically about half those of the corresponding hydrogen modes, 
consistent with the expected inverse mass scaling, as discussed more in the following section. 

\begin{table*}[t]
\centering
\caption{ Linewidths for the coherent modes of the H/metal($hkl$) systems, $\gamma_0$, as well as the vibrational linewidths obtained with the instantaneous dephasing approximation, $\bar{\gamma}$. All values are obtained from calculations using a $(4\times4\times6)$ slab with a $\Theta = 1$\,ML coverage. The reported experimental linewidths $\gamma_\text{lit}$ for those modes labelled with a $\dagger$ symbol in the table are values resulting from a fit of the Fano-type line shape given in Eq.\,\ref{eq:Fano_line} to the experimental data.  All other experimental linewidths are FWHMs of vibrational modes with a Lorentzian line shape.}
\label{tab:Linewidths_all_facets_et_isotopes}
\begin{tabular}{llcccc}
\hline
\textbf{Surface facet} & \textbf{Mode} & \textbf{isotope} &  \textbf{$\gamma_0$ / cm$^{-1}$} & \textbf{$\bar{\gamma}$ / cm$^{-1}$}  & \textbf{$\gamma_\text{lit}$} / cm$^{-1}$\\
\hline
Mo(100) & wagging   & H & 3.5  & 9  & {---} \\
        &           & D & 1.3  & 4  & {---}\\
        & sym. strech   & H & 12.0 & 29 & 65\cite{Reutt1988}\\
        &           & D & 9.6  & 18 & {---}\\
        & asym. stretch$^\dagger$ & H & 8.0 & 30 & 12\cite{Reutt1988} \\
        &           & D & 2.9  & 22 & {---}\\
\hline
W(100)  & wagging   & H & 3.2  & 2 & {---} \\
        &           & D & 1.4  & 1 & {---}\\
        & sym. strech   & H & 3.1 & 11 & 103\cite{Reutt1988}\\
        &               & D & 1.9 & 6 & 50\cite{Reutt1988}\\
        & asym. stretch$^\dagger$ & H & 4.7 & 15 & 22\cite{Reutt1988}\\
        &           & D & 1.8 & 6 & 25\cite{Reutt1988}\\
\hline
Mo(110) & wagging   & H & 13.0 & 30 & {---}\\
        &           & D & 4.5  & 13 & {---}\\
        & rocking$^\dagger$ & H & 2.1  & 20  & 35\cite{Kroger1997}\\
        &                   & D & 0.9 & 9   & 16\cite{Kroger1997}\\
        & sym. stretch  & H & 3.2  & 10 & 27\cite{Kroger1997}\\
        &           & D & 1.3  & 5  & 18\cite{Kroger1997}\\
\hline
W(110)  & wagging   & H & 3.9  & 11 & {---}\\
        &           & D & 1.7  & 5  & {---} \\
        & rocking   & H & 2.3  & 18 & {---} \\
        &           & D & 0.9  & 8  & {---} \\
        & sym. stretch  & H & 1.4 & 6 & {---}  \\
        &               & D & 0.6 & 3 & {---} \\

\end{tabular}
\end{table*}
Our computed $\gamma_0$ values of the Fano-type modes are smaller than the experimental values $\gamma_\text{lit}$. The same observation was made for static TDPT calculations of vibrational lifetimes for CO on copper surfaces.\cite{maurer16, Box2023} This finding can be easily rationalised if one recalls that the $\gamma_0$ values assume no dephasing and are therefore the lower bound of the vibrational linewidths of the adsorbates. However, the averaged linewidths $\bar{\gamma}$ are in better agreement with the experimental values as the differences are up to 18~cm$^{-1}$. Translated to vibrational lifetimes, the differences between the experimental and our computed averaged values are within $\pm$0.1\,ps. 
This confirms the assumption that all other dissipation channels for the asymmetric modes are of ancillary importance. 
The lifetimes deduced from the computed $\bar{\gamma}$ values and experimental linewidths of the modes with the Fano-line shape range from 0.1\,ps to 0.3\,ps and are significantly lower than the estimated vibrational lifetimes due to EHP-excitations for in-plane vibrations of  H on Cu(111)\cite{lamont1995} and Ni(111)\cite{Persson1991} with values of 0.7\,ps and 1\,ps, respectively. 

For those lines which exhibit a Lorentzian line shape in the experiments, we find that our computed vibrational linewidths{---}both $\gamma_0$ and $\bar{\gamma}${---}are always smaller than the experimental ones. The linewidths of the symmetric stretch mode of H/Mo(100) and H/W(100) measured by Reutt \textit{et al.}\cite{Reutt1988} are 65\,cm$^{-1}$ and 103\,cm$^{-1}$, respectively, which are significantly larger than the values we predict with our approach.  For the symmetric stretch modes of the H/metal(100) systems, EHP excitations alone cannot account for the finite linewidth. Hence, other broadening mechanisms such as adsorbate-adsorbate interactions or inhomogeneous broadening have to be taken into account, too. This is further corroborated by a pronounced temperature dependence observed for the linewidths of the symmetric stretch mode of H/W(100).\cite{Chabal1985, Reutt1988} The temperature dependence would be small if the vibrational relaxation is governed by EHP-excitations,\cite{Persson1982, langreth1995vibrational, maurer16}  which was, in fact, observed for the asymmetric stretch mode.\cite{Reutt1988} For the symmetric stretch mode of H/Mo(110), the situation is different: the difference between our calculated linewidth and the reported linewidth is similar to the difference we have seen for the asymmetric rocking mode of H/Mo(110). Combined with the fact that this linewidth was observed to be barely affected by temperature,\cite{Kroger1997}, our findings suggest that the main relaxation channel of the symmetric stretch might also be due to electron-phonon coupling.

\subsection*{Isotope Effect}

If the coupling between the electrons and the nuclei is weak, then the linewidths which are caused by EHP-excitations show an inverse dependence on the adsorbate mass $M$.\cite{Persson1982,gadzuk1984}
We quantify the kinetic isotope effect (KIE) as the H/D linewidth ratio
KIE = $\gamma_{\mathrm{H}}/\gamma_{\mathrm{D}}$ for a given mode and facet.
In the weak-coupling limit of first-order TDPT, Eq.~\ref{eq:rate_isotope_dependence}
implies $\gamma \propto 1/M$ at fixed perturbation energy, so KIE $\approx 2$ is
the canonical expectation. 
If we were to evaluate this coupling matrix for both hydrogen and deuterium at the same energy, the linewidths for the deuterium modes would be exactly half the rates for the corresponding H-vibrations. 
However, since the $\boldsymbol{\Lambda}(\hbar\omega)$ is evaluated at the respective frequencies of the H(D)-modes, the frequencies change and therefore deviations from this expected factor of two arise. 
Since deuterium vibrates at a lower frequency than hydrogen
$(\omega_{\mathrm{D}} = \omega_{\mathrm{H}}/\sqrt{2}$, in the harmonic approximation), H and D modes sample different parts of $\boldsymbol{\Lambda}(\hbar\omega)$. 
For most vibrations, we observe that the ratio of the averaged linewidths for H and D-vibrations is slightly larger than 2 which can be seen by comparing the H and D values of all modes given in Table\,\ref{tab:Linewidths_all_facets_et_isotopes}.

The isotope effect is mode and momentum-dependent. In particular,
KIE for the coherent mode is generally larger and more
dispersed than KIE for the averaged manifold. This reflects
that (a) the coherent mode samples a very narrow portion of the
electron--hole continuum and is therefore highly sensitive to the precise
perturbation energy and numerical convergence ($\kk$ mesh density and smearing width), while (b) the averaged mode integrates over backfolded
finite-$\qq$ patterns and thereby averages out some frequency- and phase–relation
sensitivities. Consequently, the KIE clusters around the weak-coupling
expectation ($\sim$2, here spanning $1.4$–$2.5$), whereas KIEs for the coherent mode
span broader values (in the range $2$–$5$).

\begin{table}[h]
\centering
\caption{Kinetic isotope effects (KIE = $\gamma_\mathrm{H}/\gamma_\mathrm{D}$) for modes where both hydrogen and deuterium linewidths are available experimentally.  
Theoretical KIEs are based on the averaged linewidths $\bar{\gamma}$ from Table~\ref{tab:Linewidths_all_facets_et_isotopes}.  Fano-type (EHP dominated) modes are indicated by~$\dagger$.}
\label{tab:KIE_comparison}
\begin{tabular}{l l c c}
\hline
\textbf{System} & \textbf{Mode} & \textbf{KIE$_\mathrm{calc}$} & \textbf{KIE$_\mathrm{exp}$}  \\
\hline
Mo(110) & Rocking$^\dagger$            & 2.2 & 2.2\cite{Kroger1997}  \\
Mo(110) & Symmetric stretch  & 2.0 & 1.5\cite{Kroger1997} \\
W(100)  & Symmetric stretch  & 1.8 & 2.1\cite{Reutt1988}  \\
W(100)  & Asymmetric stretch$^\dagger$ & 2.5 & 0.9\cite{Reutt1988}  \\
\hline
\end{tabular}
\end{table}

Since we have established that the averaged linewidths $\bar{\gamma}$ show a better agreement with the experimental linewidths, we will focus on them in the discussion of the isotope effect (Table~\ref{tab:KIE_comparison}). When comparing to experimental findings, we see that our approach correctly reproduces the isotope effect observed for the rocking mode and the symmetric stretch mode of $p(1\times1)$H(D)/Mo(110). 
Note that this is another indicator that the relaxation of the excited symmetric stretch mode of H(D)/Mo(110) is driven by the EHP-excitation mechanism.  
However, our calculations cannot capture the absent isotope effect of the lifetime for the asymmetric mode of $p(1\times1)$D/W(100) observed in the experiment.\cite{Reutt1988, Chabal1985} 
The weak isotope dependence of this mode was subject to investigation by Zhang and Langreth,\cite{Zhang1989}, who showed that the vibrational energy relaxation, when mediated by strong non-adiabatic effects, has a smaller dependence on the mass of the adsorbate. 
This interpretation would not only explain the discrepancy in the strength of the isotope effect between our calculations and the experiment for this mode, but also the slightly lower value for our computed averaged linewidths $\bar{\gamma}$ of the asymmetric stretch mode of H/W(100) in comparison to the experimental lifetime. 

\subsection*{Influence of the hydrogen coverage}
\begin{figure}
    \centering
    \includegraphics[width=3.3in]{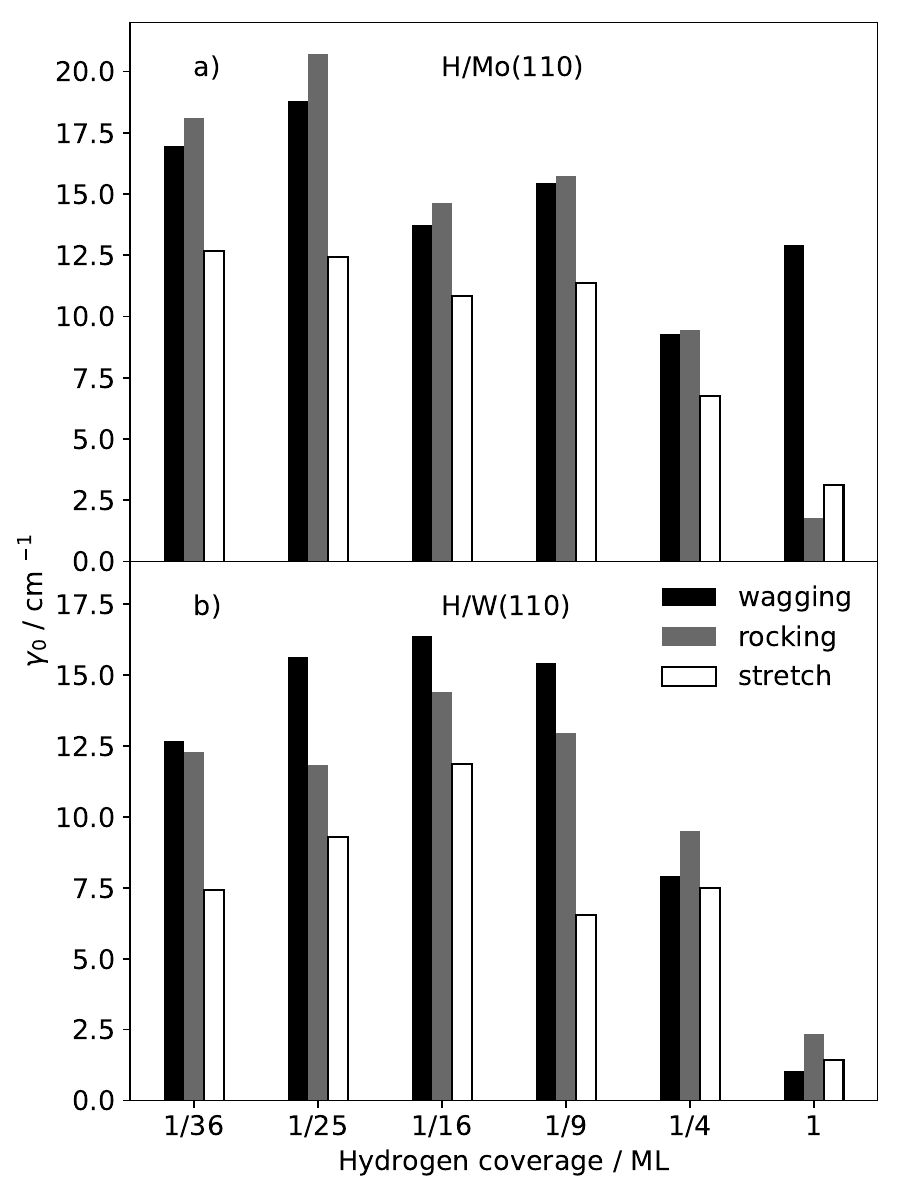}
    \caption{ Electron-phonon-coupling-induced vibrational linewidths of the three different collective modes of hydrogen adsorbed on Mo(110), shown in panel a), and W(110), shown in panel b)  as a function of hydrogen coverage $\Theta$. The $1/n^2$ coverages were realised by placing one hydrogen on the three-fold side of a $p(n\times n)$ surface slab. }
    \label{fig:Relaxation_rates}
\end{figure}

Thus far, we have discussed H/metal systems with a full surface coverage $\Theta=1$\,ML.
We will investigate coverage dependence for the bcc(110) surfaces as Mo(100) and W(100) undergo different surface reconstructions at different hydrogen coverages.\cite{Prybyla1987}  The relaxation rates for the coherent modes, $\gamma_0$, of hydrogen on both bcc metal(110) surfaces are shown in Figure\,\ref{fig:Relaxation_rates}. 

The EHP-induced linewidths increase markedly as the hydrogen coverage decreases,
with enhancements of up to a factor of eight across the different substrates and
vibrational branches. 
Although we do not compute the density of states
explicitly here, high adsorbate coverages are expected to alter the surface
electronic structure and therefore the density of available electronic
excitations that contribute to Eq.~\ref{eq:LambdaGamma}. Such coverage-related
changes in the local electronic environment offer a plausible explanation for
the weaker relaxation of the coherent modes at full monolayer coverage.
For H/Mo(110) systems, the relaxation rates for the vibrations start to plateau once the coverage has reached 1/9\,ML. For H/W(110), the linewidths rise until they reach  1/16\,ML coverage and then show a trend of decline with decreasing coverage.   
Furthermore, we also observe that the strength of the coverage dependence of the relaxation rates is also dependent on the mode. The wagging mode of H/Mo(110) shows the weakest dependence on coverage among all modes, whereas the corresponding mode of H/W(110) exhibits a noticeably stronger one. We attribute this difference to a more pronounced coverage dependence of the non-adiabatic couplings in tungsten than in molybdenum, as discussed in more detail below.
The symmetric stretch mode of hydrogen on both metals shows a smaller dependence in comparison to the rocking mode. Both modes have similar linewidths for $\Theta=1$\,ML but when the coverage is decreased, the rocking mode increases more strongly than the asymmetric stretch mode. 



In Table\,\ref{tab:Friction_coverage_screen}, the entries of the non-adiabatic coupling matrix in the zero-frequency limit, $\boldsymbol{\Lambda}(\hbar \omega \rightarrow 0)$, are presented. In the zero-frequency limit, this quantity can be interpreted as Markovian electronic friction tensor.\cite{head-gordon95, maurer16}
\begin{table*}
    \centering
    \caption{Entries of the friction tensor $\boldsymbol{\Lambda}(0)$, as a function of hydrogen coverage on Mo(110) and W(110).The last entry, $\Lambda_\text{iso}$, stands for the isotropic friction, which serves as a measure for the magnitude of electronic friction which the adsorbate experiences.  The different coverages were realised by placing one hydrogen on the three-fold side of a $p(n\times n)$ surface slab. All elements with values below 0.01\,ps$^{-1}$ are neglected, which is why $\Lambda_{xy}$ and $\Lambda_{xz}$ are not shown in the table.}
    \label{tab:Friction_coverage_screen}
    \begin{tabular}{l l c c c c c c}
    \hline \hline
     System    & Cell size  &  $\Theta$ / ML & $\Lambda_{xx}$ / ps$^{-1}$ &  $\Lambda_{yy}$ / ps$^{-1}$ &  $\Lambda_{zz}$ / ps$^{-1}$ & $\Lambda_{yz}$ / ps$^{-1}$ & $\Lambda_\text{iso}$ /  ps$^{-1}$ \\ \hline
H/Mo(110)  & $p(1 \times 1)$ & 1     & 1.64 & 0.31 & 0.39 & 0.04 & 0.78 \\
           & $p(2 \times 2)$ & 1/4   & 1.17 & 1.04 & 0.79 & 0.26 & 1.00 \\
           & $p(3 \times 3)$ & 1/9   & 1.63 & 1.61 & 1.13 & 0.28 & 1.46 \\
           & $p(4 \times 4)$ & 1/16  & 1.51 & 1.47 & 1.18 & 0.26 & 1.38 \\
           & $p(5 \times 5)$ & 1/25  & 1.87 & 1.96 & 1.22 & 0.34 & 1.68 \\
           & $p(6 \times 6)$ & 1/36  & 1.79 & 1.79 & 1.31 & 0.32 & 1.63 \\
           &                 &       &       &       &       &       &       \\
H/W(110)   & $p(1 \times 1)$ & 1     & 0.29 & 0.23 & 0.27 & 0.09 & 0.26 \\
           & $p(2 \times 2)$ & 1/4   & 0.87 & 0.84 & 0.71 & 0.23 & 0.81 \\
           & $p(3 \times 3)$ & 1/9   & 1.64 & 1.25 & 0.93 & 0.38 & 1.27 \\
           & $p(4 \times 4)$ & 1/16  & 1.66 & 1.45 & 1.09 & 0.15 & 1.40 \\
           & $p(5 \times 5)$ & 1/25  & 1.69 & 1.41 & 0.89 & 0.23 & 1.33 \\
           & $p(6 \times 6)$ & 1/36  & 1.44 & 1.21 & 0.95 & 0.31 & 1.20 \\
 \hline

    \end{tabular}    
\end{table*}
The $xx$-component of the friction tensor, $\Lambda_{xx}$, differs significantly between the tungsten and molybdenum surfaces. For Mo(110), this entity is insensitive to changes in the coverage, whereas it shows a strong coverage dependence in the case of W(110). This is the reason why the calculated $\gamma_0$ values for the wagging mode of H/Mo(110) do not increase when hydrogen coverage at the surface drops. All other entries of the friction tensor of both systems increase with decreasing coverage and start to plateau around a coverage of $\Theta=1/9$~ML. Furthermore, the diagonal elements of the friction tensor are clearly different for H/Mo(110) at $\Theta=1$~ML. The friction tensor for H/W(110) at this coverage, on the other hand, is nearly isotropic. At low coverage, on the other hand, the $\Lambda_{zz}$ component is the smallest friction entry on the diagonal, whereas $\Lambda_{xx}$ and $\Lambda_{yy}$ are comparatively similar. Hence, the friction tensors for hydrogen atoms close to the surfaces are anisotropic.
For both systems, we find that the off-diagonal elements are small compared to the diagonal elements. For both H/Mo(110) and H/W(110), the $yz$-component of the friction tensor, $\Lambda_{yz}$, is the only significant off-diagonal element which reflects the geometry of the surface as the $yz$ direction points directly towards a metal atom. Whilst at high coverages, $\Lambda_{yz}$ is almost insignificant, it is almost three or four times smaller than $\Lambda_{zz}$ at small coverages and thence we conclude that off-diagonals cannot be omitted, per se, if one wishes to study hydrogen diffusion on metal surfaces or diffusion into the interior of the substrate.

Our findings have implications for molecular dynamics with electronic friction (MDEF): an established technique in theoretical surface science for including non-adiabatic effects into molecular dynamics at metal surfaces.\cite{li_wahnstrom92a, head-gordon95, auerbach21} This technique has proven highly successful in describing the energy transfer dynamics observed in H atom scattering experiments on pristine metal surfaces\cite{buenermann15, Dorenkamp18, hertl22b} and hot H atom dissipation,\cite{blanco14, Galparsoro18} but was shown to fail for adsorbate-covered surfaces.\cite{lecroart21} Note that in these studies, friction was treated as a single coefficient dependent on the background electronic density of the pristine metal surface, an approach commonly referred to as the local density friction approximation (LDFA).\cite{li_wahnstrom92a, li_wahnstrom92b, juaristi08} The LDFA and TDPT approaches give similar results for hydrogen dynamics at metal surfaces provided that the coverage is low.\cite{spiering2018, spiering2019, box2024, stark2025} In light of the coverage dependence of the friction tensor and the agreement of the here computed linewidths with spectroscopic experiments, we conclude this agreement between LDFA and TDPT might not generally hold true at high coverages. Inelastic and reactive hydrogen atom scattering from densely hydrogen-covered tungsten surfaces was modelled with LDFA-MDEF, and in those studies, the couplings between the electrons and the nuclei were found to be the main dissipation channel, whereas coupling to vibrations was weaker.\cite{martin-barrios2021, martin-barrios2024, Galparsoro2025} Close to the surface, LDFA-based friction coefficients for hydrogen were shown to have a value of $\sim$6\,ps$^{-1}$ for a variety of transition metal surfaces,\cite{juaristi08, janke15, box2024} including molybdenum and tungsten.\cite{ hertl22, martin-barrios2022} This might also affect computed laser-induced desorption simulations on densely hydrogen-covered Ru(0001).\cite{juaristi2017}
We introduce the isotropic friction tensor
\begin{equation}
    \Lambda_\text{iso} = (\Lambda_{xx} + \Lambda_{yy} + \Lambda_{zz})/3
\end{equation}
to compare the magnitude of TDPT friction with the isotropic LDFA approximation. As one can infer from the last column of Table\,\ref{tab:Friction_coverage_screen}, the $\Lambda_\text{iso}$ value for H/W(110), when densely covered, is about 20 times smaller than its LDFA counterpart. On the other hand, at low coverages, LDFA-based friction coefficients with ~6\,ps$^{-1}$ for H/Mo(110) and H/W(110) are still 4{--}5 times larger than the values predicted with TDPT, but under these conditions, the off-diagonal terms make a notable contribution, which are also omitted in LDFA. Hence, we advise more caution when dealing with non-adiabatic effects in the friction picture for dynamics on densely-covered surfaces as the reported dominant role of EHP excitations in the energy dissipation dynamics of hydrogen at metal surfaces with high coverages may, in part, reflect an overestimation of the strength of non-adiabatic effects; other contributions arising from adsorbate–adsorbate interactions and second order electron-phonon coupling could also play a more significant role than previously assumed.\cite{Novko2018} 
Since metal surfaces in technological processes are not pristine, our theoretical framework to compute coverage dependent friction tensors might therefore contribute to more accurate modelling of sticking coefficients of hydrogen on adsorbate covered surfaces{---}an important elementary step for modelling plasma dynamics in a fusion reactor.\cite{causey2002}

\section*{Conclusion}
In this work, we carried out first principles electron-phonon coupling calculations for hydrogen adsorbed on the (100) and (110) surfaces of molybdenum and tungsten, enabling us to determine EHP-induced vibrational linewidths for all adsorbate modes. The collective vibrational linewidths are generally smaller than the corresponding averaged values, reaching an order of magnitude difference in some cases. For the modes that exhibit Fano line shapes in experiment, the collective-mode linewidths alone do not reproduce the measured widths, with the possible exception of the asymmetric stretch of hydrogen on molybdenum(100).

Under the assumption of instantaneous dephasing, our calculated $\mathbf{q}$-averaged linewidths show good overall agreement with the experimentally fitted FWHMs for Fano-type modes, consistent with rapid electron-mediated dephasing of the collective vibration. The averaged linewidths typically agree within a factor of two with experiment for all modes that display Fano line shapes. A notable exception is the asymmetric stretch of hydrogen on molybdenum(100), where our value is roughly twice as large as the experiment. 

For Lorentzian line shapes, the situation is more varied. The symmetric stretch modes of hydrogen and deuterium on tungsten(100) are underpredicted by about an order of magnitude, indicating that other dissipation mechanisms may be dominant, such as phonon-phonon coupling. Other recent studies have compared the magnitude of these effects.~\cite{Hoermann2025} 
In contrast, the symmetric stretch modes of hydrogen on molybdenum(100) and molybdenum(110) yield averaged linewidths that lie within a factor of two of the experimental values. For the symmetric stretch on molybdenum(110), the similarity between its deviation and that of the asymmetric rocking mode, combined with its negligible temperature dependence reported in experiment,\cite{Kroger1997}, suggests that its relaxation pathway may also be dominated by coupling with EHP excitations.

By explicitly calculating the coverage-dependent linewidths and electronic friction tensors for hydrogen on molybdenum(110) and tungsten(110), we find a significant and systematic coverage dependence in both quantities. The coherent vibrational linewidths remain relatively consistent at low coverages but decrease noticeably at the highest coverages ($\Theta = 1/4$ and $1$~ML). A notable exception is the wagging mode on molybdenum(110) at one monolayer, which shows an unusually large linewidth caused by strong mode anisotropy. This trend is closely reflected in the friction tensor. At full monolayer coverage, the hydrogen--molybdenum(110) system becomes strongly anisotropic, whereas hydrogen--tungsten(110) shows a similar reduction in friction at high coverage but retains a comparatively isotropic character in its diagonal elements. At sparse coverages, among all directions, motion along the surface normal experiences significantly reduced electronic friction than parallel to the surface, indicating pronounced anisotropy.

These findings have direct implications for molecular dynamics with electronic friction. Approaches that rely on isotropic electronic friction approximations, such as the LDFA approach, may be suitable at low coverages, but likely overestimate electronically non-adiabatic effects when applied to densely covered surfaces.

\section*{Appendix}
We checked the influence of the chosen exchange-correlation functional on the friction tensor $\bm{\Lambda}(0)$ at the level of the local density approximation and different functionals at generalised gradient approximation. We find that the choice of the exchange-correlation functional does not have significant impact on the friction tensor{---}see Table\,\ref{tab:xc-functional}. 
\begin{table}[h]
\centering
\caption{Elements of Cartesian friction tensor $\bm{\Lambda}(0)$ for the primitive $p(1\times1)$H/Mo(110) cell as a function of different DFT functionals. All elements with values below 0.01\,ps$^{-1}$ are neglected, which is why $\Lambda_{xy}$ and $\Lambda_{xz}$ are not shown in the table.}
\label{tab:xc-functional}
\begin{tabular}{l c c c c }
\hline
\textbf{Functional} & $\Lambda_{xx}$/ps$^{-1}$ & $\Lambda_{yy}$/ps$^{-1}$ & $\Lambda_{zz}$/ps$^{-1}$  & $\Lambda_{yz}$/ps$^{-1}$\\
\hline
LDA & 1.68 & 0.31 & 0.42 & 0.05\\
PBE & 1.64 & 0.31 & 0.39 & 0.04\\
PW91 & 1.72& 0.32 & 0.38 & 0.04\\
RPBE & 1.63& 0.31 & 0.38 & 0.04\\
\hline
\end{tabular}
\end{table}

\section*{Author contributions}
N.H. conceptualised the project. R.J.M. supervised the project. N.H. performed the calculations and analysed and curated the data. N.H., C.L.B., and R.J.M.  discussed and interpreted the findings. N.H. and C.L.B. wrote the original draft. N.H., C.L.B., and R.J.M. edited and reviewed the manuscript.

\section*{Data availability statement}

All linewidth and electronic friction calculations discussed in this manuscript are available within the NOMAD
repository (\href{https://dx.doi.org/10.17172/NOMAD/2025.11.17-1}{10.17172/NOMAD/2025.11.17-1)}.


\section*{Conflicts of interest}
There are no conflicts to declare.

\section*{Acknowledgments}
Funding is acknowledged from the UKRI Future Leaders Fellowship programme (MR/X023109/1), a UKRI frontier research grant (EP/X014088/1), and an MSCA postdoctoral fellowship (EP/Z001498/1). High-performance computing resources were provided via the Scientific Computing Research Technology Platform of the University of Warwick, the EPSRC-funded Materials Chemistry Consortium (EP/R029431/1, EP/X035859/1), and the UK Car-Parrinello consortium (EP/X035891/1) for the ARCHER2 UK National Supercomputing Service, and the EPSRC-funded HPC Midlands+ computing centre for access to Sulis (EP/P020232/1).







\bibliography{Literature} 

\bibliographystyle{rsc} 



\end{document}